\documentclass[twocolumn]{aastex631}

\usepackage{amsmath}
\usepackage{booktabs}
\usepackage{graphicx}
\usepackage[normalem]{ulem}

\newcommand{\msol}{$\mathrm{M}_\odot$}
\newcommand{\msolyr}{$\mathrm{M}_\odot$ yr$^{-1}$}
\newcommand{\kms}{km s$^{-1}$}
\newcommand{\mdoti}{$\dot{M}_\mathrm{ion}$}
\newcommand{\mdotn}{$\dot{M}_\mathrm{neut}$}

\newcommand{\hi}{\ion{H}{1}}

\newcommand{\cii}{\ion{C}{2}}
\newcommand{\civ}{\ion{C}{4}}
\newcommand{\siii}{\ion{Si}{2}}
\newcommand{\siiii}{\ion{Si}{3}}

\newcommand{\oi}{\ion{O}{1}}
\newcommand{\ovi}{\ion{O}{6}}

\newcommand{\rev}[2]{#2}

\defcitealias{lucchini24}{Paper I}

\begin{document}

\title{Invisible Accretion: Ionized Envelopes of \rev{High-velocity Clouds}{TNG50 HVCs} can Sustain \rev{Galactic }{}Star Formation}

\author[0000-0001-9982-0241]{Scott Lucchini}
\affiliation{Center for Astrophysics $|$ Harvard \& Smithsonian, 60 Garden Street, Cambridge, MA 02138, USA}

\correspondingauthor{Scott Lucchini}
\email{scott.lucchini@cfa.harvard.edu}

\author[0000-0002-6800-5778]{Jiwon Jesse Han}
\affiliation{Center for Astrophysics $|$ Harvard \& Smithsonian, 60 Garden Street, Cambridge, MA 02138, USA}

\author[0000-0001-6950-1629]{Lars Hernquist}
\affiliation{Center for Astrophysics $|$ Harvard \& Smithsonian, 60 Garden Street, Cambridge, MA 02138, USA}

\author[0000-0002-1590-8551]{Charlie Conroy}
\affiliation{Center for Astrophysics $|$ Harvard \& Smithsonian, 60 Garden Street, Cambridge, MA 02138, USA}

\author[0000-0003-0724-4115]{Andrew J. Fox}
\affiliation{AURA for ESA, Space Telescope Science Institute, 3700 San Martin Drive, Baltimore, MD 21218, USA}
 
\begin{abstract}

Galactic high-velocity clouds (HVCs) are known to be complex, multiphase systems consisting of \rev{neutral hydrogen visible in 21~cm emission maps, along with additional ionized material and in some cases, cooler molecular gas}{neutral and/or ionized gas moving at high velocities relative to the rotation of the disk}. In this work, we investigate Milky Way-like galaxies from the TNG50 simulation to characterize the properties, morphology, and accretion rates of the warm and hot ionized material comoving with neutral HVCs visible in \ion{H}{1}. We find that the ionized gas forms an envelope around the neutral material, and in most cases (73\% of the HVCs) it is prolate in morphology. We also find that the ionized mass is $\sim6$ times greater than the neutral mass, which leads to significantly more accretion possible from the ionized gas (\mdoti) than the neutral gas (\mdotn), consistent with estimates made from observations of our own Galaxy. We investigate the accretion rates from both phases of HVCs around 47 Milky Way-like galaxies and find that \mdoti\ scales with \mdotn, and both scale with the star formation rate of the galaxy. Finally, we find that, on average, \mdoti\ could account for 81\% of the galactic star formation rate (assuming the material can sufficiently cool and condense), while \mdotn\ can only balance 11\%. Thus, the diffuse, ionized, high-velocity circumgalactic medium plays a defining role in the evolution and growth of galaxies at low redshift.

\end{abstract}


\section{Introduction} \label{sec:intro}

The accretion of gaseous material onto galactic disks is required for sustained star formation and galaxy growth. However, where this material comes from and what form it takes is still largely unknown. We do know that galaxies are, in general, surrounded by a multi-phase gaseous halo (the circumgalactic medium, or CGM) and all material that eventually falls onto the disk must pass through this region \citep{spitzer56,binney77,chevalier79,tumlinson17}. Thus to understand how galaxies continually replenish their fuel for star formation, we must understand the properties and dynamics of the CGM.

Our own Milky Way Galaxy (MW) offers an excellent test-bed for studies of the CGM, and therefore galaxy evolution, as we are able to get a much more complete picture of its current state and properties due to its proximity and ubiquity. 21 cm \ion{H}{1} emission maps of \rev{}{selected sources as well as} the entire sky have been available for decades, and allow us to image the morphology of neutral gas structures in the inner CGM \citep{murphy95,kalberla05,hi4pi,westmeier18}. To isolate these distant structures from the overwhelming emission from the disk and local environment, we have restricted this study to objects that are moving at anomalous velocities relative to the rotation of the disk. These ``high-velocity clouds'' (HVCs) give us a view of the neutral CGM of our own Galaxy \citep{vanwoerden57,muller63,wakker97,benjamin04,putman12}.

Alternatively, we can detect gas via absorption \citep{wakker01,collins07,collins09,shull09,lehner11,lehner12,fox14,richter17,fox19}. Taking a spectrum of a distant quasar gives us an immense amount of information about the intervening, low-density material due to the fact that this material can absorb specific wavelengths of the quasar's light, leaving an imprint in the resulting observed spectrum. By quantifying the profiles of the absorption lines, we can learn about the composition, density, and velocity of the intervening gas.
By associating the velocities of the various ions with the \ion{H}{1} emission maps, we can obtain a multiphase view of the Milky Way's CGM structures.

While \ion{H}{1} maps can give us the morphologies and structure of HVCs, absorption spectroscopy can provide information on their composition, ionization state, dust properties, and distances. Depending on the instrument, this technique can be sensitive to a variety of different absorption lines, such as \ion{H}{1} 21~cm and molecular lines in the radio (e.g. VLA, GBT, Parkes, ALMA; \citealt{westmeier18,steffes24}), H$_2$, \ion{C}{2}, \ion{C}{4}, \ion{Si}{2}, \ion{Si}{3}, \ion{Si}{4}, and \ion{O}{6} in the ultraviolet (e.g. HST COS\rev{}{, FUSE}; \citealt{richter99,sembach03,fox19,french21}), and H$\alpha$, Ca~H and K, \ion{Mg}{2}, and Na~D\rev{, and molecular lines}{} in the optical and infrared (e.g. Keck, VLT; \citealt{tufte98,wakker00,mishra25}). These observations not only give us information about how highly ionized and metal rich the gas is, but by comparing the abundances of these different elements, we can learn about the physical properties of the material such as how much dust depletion there is \citep{lu98,gibson00,fox23}.
Finally, by using MW halo stars instead of quasars as backlights, we can obtain distance limits for the detected gas \citep{smoker11,lehner22}. If absorption is seen (not seen), the gas must be closer (farther) than the background star used. \rev{}{This can be used to construct a population-based HVC covering fraction as a function of distance (as in \citealt{lehner22}), or to constrain individual HVCs with sets of stars at different distances projected close together on the sky (as in \citealt{smoker11}).}

While ionized material has long been seen in quasar absorption spectra \citep{sembach03,fox06,richter09}, its association with the Galaxy remained uncertain until its detection towards Galactic halo stars by \citet{lehner11}, limiting its distance to 15~kpc from the Sun. Furthermore, this ionized material is theoretically required to fuel star formation \citep{bauermeister10}. Thus the identification of ionized HVCs (with N(\ion{H}{2}) $>$ N(\ion{H}{1})) connected our galaxy with extragalactic sources and our picture of cosmological galaxy growth.

Estimates from the data we do have suggest that the amount of ionized material vastly outweighs the visible neutral HVCs by a factor of $\gtrsim$~6:1 \citep{shull09,lehner11,fox19}. The Magellanic Stream dominates the high velocity material in both neutral and ionized gas which could accrete onto the Galaxy at a rate of $1.9-3.4$~\msolyr\ \citep{fox14}. \rev{}{However, the fate of this material is uncertain as it may never reach the disk due to heating or turbulence as it passes through the MW CGM \citep[e.g.][]{murali00}.} We exclude this material from our study due to the serendipitous nature of the current interactions between the MW and the Magellanic Clouds. The remainder of the high velocity sky is estimated to have an inflow rate of 0.08~\msolyr\ from \ion{H}{1} radio maps \citep{putman12}, while latest values including ionized material are $\gtrsim0.53\pm0.23$~\msolyr\ \citep{fox19}. Compared against estimates of the Galactic star formation rate ($0.7-1.9$~\msolyr; \citealt{robitaille10,chomiuk11,licquia15}), \rev{the ionized material could provide the dominant accretion source}{this ``invisible accretion'' from ionized material not seen in emission maps of our CGM could provide the dominant source of star-forming fuel to our disk}. However, \rev{there are still many unknowns due to observational constraints.}{obtaining the full 3D motions and morphologies of these clouds remains out of reach observationally.}

Cosmological galaxy simulations allow us to directly probe the ionized components of HVC analogs to quantify their properties and answer the question of where a galaxy's star formation fuel comes from. In this paper, we use the TNG50 simulation from the IllustrisTNG simulation suite \citep{nelson19b,nelson19,pillepich19} to make mock absorption line observations of Milky Way-like galaxies to compare with \rev{the data}{observational covering fractions and mass flow rates}. 
We then characterize the properties of the ionized HVCs and investigate the masses and accretion rates of the various phases in relation to galactic star formation.
In Section~\ref{sec:methods}, we outline our methods for observing and identifying the HVC ionized envelopes. Section~\ref{sec:results} describes our main results, and we discuss and conclude in Sections~\ref{sec:discussion} and \ref{sec:conclusions}.

\section{Methods} \label{sec:methods}

IllustrisTNG is a suite of cosmological box simulations including TNG50-1, a run with a box size of 50~Mpc and baryonic resolution of $8.5\times10^{4}$~\msol\ \citep{nelson19b,nelson19,pillepich19}. The data products from this massive project have been made available to the public\footnote{\url{http://www.tng-project.org}} and with these results, we have investigated several MW-like galaxies from this simulation box. Initially, we analyzed the same galaxy that we used in our first paper (\citealt{lucchini24}, hereafter \citetalias{lucchini24}), galaxy subhalo ID 537941. In \citetalias{lucchini24} we characterized the origins of HVC analogs detected around this galaxy. To do this, we identified all the contiguous cold ($T<10^{4.5}$~K) gas cells within 200~ckpc of the galactic disk. From this set of cold clouds, we compared the mean line of sight velocity per cloud with the range in possible values for galactic rotation along the direction toward the cloud. If the cloud's velocity deviated by more than 70~km~s$^{-1}$ from galactic rotation, it was deemed an HVC analog. For a full description of this method, see Sections~2.2 and 2.4 in \citetalias{lucchini24}. Here, we use the same sample of HVCs, and follow the same technique to identify HVCs around other galaxies in our sample (see Section~\ref{sec:population}).

\rev{}{Throughout this paper, we refer to our identified cold clouds as our HVC analogs (corresponding to the \ion{H}{1} HVCs visible in the emission around the MW; see \citetalias{lucchini24}), and we refer to the warm-hot, ionized material comoving with the cold clouds as the ionized HVCs or the ionized envelopes. We acknowledge that in the observations of the MW's CGM, there are many instances of ionized high velocity material not associated with any neutral feature seen in emission. We do not explicitly study these structures in this work, however we do see evidence in agreement with the observations based on the covering fractions of various ions discussed below.}

\subsection{Mock Absorption Spectra}

In order to \rev{ensure that our simulation reproduced realistic, multiphase HVCs,}{directly compare with observations,} we performed mock absorption line spectroscopy \rev{for each identified cloud}{through our simulated CGM}. \rev{}{The main goal is to qualitatively determine if the simulations contained comoving, multiphase gas along any given sightline. Below, we also calculate covering fractions and mass flow rates directly from the mock spectra as well.}

For this, we used \textit{Trident}\footnote{\url{https://github.com/trident-project/trident}} \citep{hummels17}. \textit{Trident} is built upon \textit{yt}\footnote{\url{https://yt-project.org}} \citep{turk11} and utilizes pre-computed \textsc{cloudy} ionization tables to track chemical compositions within a simulation. \rev{By specifying a path it can generate absorption spectra and column densities for many different elements and ions.}{The IllustrisTNG simulations do self-consistently track the evolution of 10 elements injected via stellar feedback. The \textit{Trident} code combines these metal mass fractions, as well as gas cell temperatures and densities, with the \textsc{cloudy} tables to determine population fractions for specific ionic states. In addition to calculating these properties simulation-wide, \textit{Trident} can also build up absorption spectroscopy profiles for specified sightlines.} We are using \textit{yt} version 4.4 and the high resolution tabulated UV background from \citet{haardt12} including self-shielding, and we explicitly track \ion{H}{1}, \ion{C}{2}, \ion{C}{4}, \ion{Si}{2}, \ion{Si}{3}\rev{}{, \ion{O}{1}, and \ion{O}{6}}.

\rev{}{These mock spectra are created with a velocity resolution of 6.5~\kms, signal to noise ratio of 25, and convolved with a line-spread function to match the E140M grating of the \textit{Space Telescope Imaging Spectrograph} (STIS) on the \textit{Hubble Space Telescope}.}

\rev{}{In addition to investigating individual spectra, we performed a statistical analysis of the simulated CGM using 300 random sightlines. Following \citet{richter17} and \citet{fox19}, we calculated the covering fraction and mass flow rates purely from these 300 spectra. We did not perform any Voigt profile fitting, we simply restricted our simulation to only the high-velocity material and summed up the column density and calculated a column density weighted mean velocity along a line of sight. For each sightline, we assume the gas is at a distance of 12~kpc, as in \citet{fox19}. We use \ion{Si}{2} and \ion{Si}{3} to compute the total hydrogen mass flow rates}
\begin{equation} \label{eq:nh}
    N_\mathrm{H} = \frac{N(\mathrm{Si\ II)+N(\mathrm{Si\ III)}}}{Z(\mathrm{Si}/\mathrm{H})_\odot}
\end{equation}
\rev{}{with $Z$ as the overall metallicity, and (Si/H)$_\odot=3.24\times10^{-5}$ as the solar silicon abundance. From a covering fraction ($f_c$), mean column density, mean velocity, and distance ($d$), we can compute the mass flow rates:}
\begin{equation} \label{eq:flowrate}
    dM/dt = \left(1.4m_\mathrm{H} f_c\langle N_\mathrm{H} \rangle 4\pi d^2\right)\times \langle v\rangle/d
\end{equation}
\rev{}{which can be computed for the inflowing or outflowing material separately. Uncertainties on the covering fractions were calculated using the Wilson score interval ($\sigma_{f_c}=\sqrt{f_c(1-f_c)/n}$, where $n$ is the number of sightlines), and uncertainties on the means are standard errors \citep{fox19}.}

\begin{figure}
    \centering
    \includegraphics[width=0.7\columnwidth]{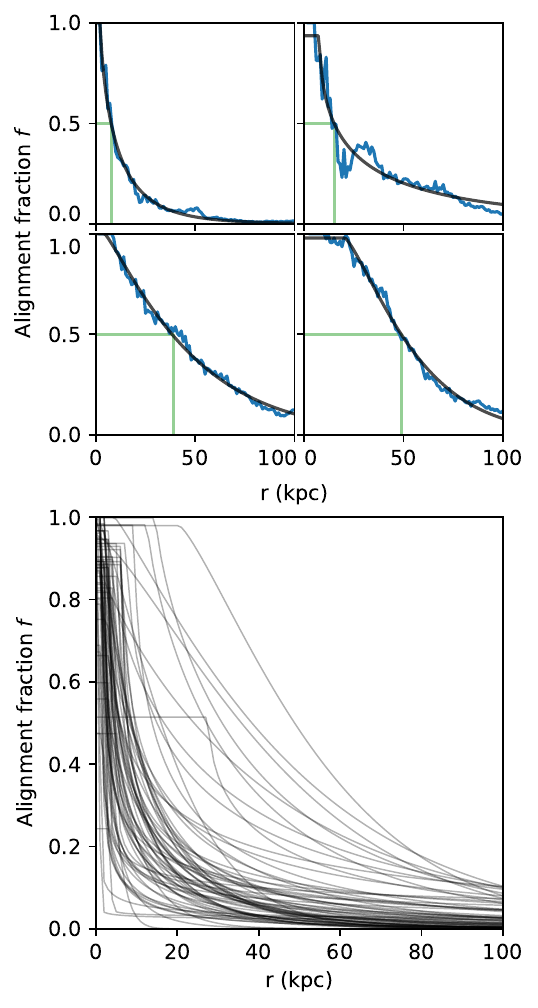}
    \caption{The fraction of hot ($T>10^{4.5}$ K) gas cells within a given spherical shell around an HVC that have velocities aligned with the mean HVC velocity. The top panels show the alignment fractions for four HVCs as measured in the simulation in blue. The black lines are fits using a cored exponential profile (see Equation~\ref{eq:alignfrac}). The green lines denote the half-size for these clouds$-$the radius at which the fraction of gas cells with velocities aligned with the HVC drops below 50\%. The bottom panel shows the exponential fits for all 70 HVCs.}
    \label{fig:alignfrac}
\end{figure}

\subsection{Ionized Envelope Identification} \label{sec:ion}

We then used the simulations to isolate the 3D position of this ionized material in relation to our \rev{}{neutral} HVC analogs. Assuming the comoving ionized gas is surrounding the HVC, we aim to determine its structure in the following way.
We define the alignment factor, $dv_i$ for a given cell $i$, as the normalized difference between the velocity vector of that cell, and the mean velocity of the HVC: $dv_i\equiv \vert v_i - v_\mathrm{HVC}\vert/\vert v_\mathrm{HVC}\vert$.

We then define the alignment fraction, $f$, as a function of radial distance away from the mean position of the HVC, in order to define an approximate size of the ionized envelope.
$f$ is the fraction of gas cells at a given radius whose velocity vector is aligned with the velocity of the HVC; specifically, $dv_i \leq 0.5$. This function is then fit with a cored exponential profile,
\begin{equation} \label{eq:alignfrac}
    f(r) = 
    \begin{cases}
        f_\mathrm{max} & r\leq r_c\\
        f_\mathrm{max} e^{-(r-r_c)^\alpha/\tau} & r>r_c
    \end{cases}
\end{equation}
with the core radius, $r_c$, max fraction, $f_\mathrm{max}$, exponential shape, $\alpha$, and scale, $\tau$, as free parameters for the fit.

To quantify the size of these ionized envelopes, we define the ``half-size'' as the radial distance at which $f=0.5$. Throughout the rest of this paper, we define the ionized envelope around each HVC as the gas with $T>10^{4.5}$~K weighted by the alignment with the HVC velocity, $\max(0,1-dv_i)$, within twice the half-size radius.

Figure~\ref{fig:alignfrac} shows the radial distributions of the alignment fraction for the HVCs. The top panels show four examples with the measured alignment fractions in blue, the fit in black, and the half-size in green. The bottom panel shows the fits for all 70 HVCs.

We also calculate the accretion rates due to the HVCs and their ionized envelopes. For each neutral HVC, we sum up the mass of its constituent cells, multiply by the radial component of the mean HVC velocity, and divide by the mean galactocentric distance to the HVC: \rev{}{$\dot{M}_\mathrm{neut}=M_\mathrm{neut}\times v_{r,\mathrm{neut}} / r_\mathrm{neut}$}. The total accretion rate due to all HVCs is therefore the sum of this quantity for all infalling HVCs.

The corresponding calculation for the ionized envelopes is slightly different so as to ensure we don't double-count any material. For each gas cell in the simulation, we associate a scalar field, $w$, initialized to zero. Then for each HVC we determine the alignment fraction for all the cells within twice the half-radius, calculate the weights as above, $\max(0,1-dv_i)$, and add those values to the scalar fields associated to each cell. We can then sum up the accretion rate contributions for each gas cell weighted by the alignment fraction (with a maximum weight value of 1): $\dot{M}_\mathrm{ion}=\sum_i M_i \min(1,w_i) v_{r,i} / r_i$. By restricting this sum to only the infalling material, we arrive at a total accretion rate onto the galaxy due to the ionized envelopes of the HVCs.

In order to determine the morphology of the envelopes, we follow a similar procedure as outlined in \citetalias{lucchini24} to identify cold clouds.
Starting with the Voronoi cells that comprise the HVC, we walk through the Voronoi neighbor tree and include any cell with $dv_i$ less than some cutoff value. 
Once no further neighbors satisfy the criteria, we end the search, resulting in a connected set of Voronoi cells that constitute the comoving envelope of the HVC. In order to obtain the ionized envelope specifically, we apply an additional mask to the connected envelope cells such that their temperatures are greater than $10^{4.5}$~K. We begin with a $dv_i$ cutoff value of 0.5 (as in the alignment fraction calculation), however in some cases, this leads to \rev{ionized envelopes extending across the entire galaxy due to thin connections between groupings of low $dv_i$ cells}{unrealistically large ionized envelopes due to random alignment of ambient CGM gas cells}. These extended envelopes clearly include material that is not associated with the individual HVC, so we sequentially step down in $dv_i$ cutoff value by steps of 0.05 until we find an envelope with fewer than 2,000 cells. We have tried max envelope sizes of 1,000 and 3,000 as well, and we find consistent results with 2,000 cells providing the best agreement with the half-size calculations.

\subsection{Impact of Resolution}

With all simulations, trade-offs must be made between resolution and scale. TNG50 provides an excellent balance between these two with gas cell resolution of $8.5\times10^4$~\msol\ and a cosmological volume of 50~Mpc$^3$ \citep{nelson19}. This equates to a spatial resolution less than 1~kpc within 13~kpc off the plane of the galaxy (see Figure~1 in \citetalias{lucchini24}) while also providing multiple MW-like galaxies that we can study (198 in \citealt{pillepich24}; 61 in \citealt{semenov24}; 47 used in this work). However, resolution remains a limiting factor in resolving the CGM due to its immense volume and low densities.

\rev{}{It has been shown that the specific properties of mock absorption spectra depend greatly on resolution \citep{foggie}. At lower resolutions, each cell is contributing an outsized feature in the resultant spectrum. Higher resolutions allow many smaller cells to build up more complex and realistic spectral profiles. However, for this work we are not analyzing the spectra themselves or performing detailed Voigt fits. We are simply using them as a qualitative metric for whether we can reproduce the comoving, multiphase gas that is seen in observations. And it has also been shown that the total gas mass, covering fractions, and large-scale structures are converged at relatively low resolutions \citep{foggie,gible,augustin25}. Thus discrepancies in these properties when compared to observations are likely a result of missing physical processes such as radiative transfer, thermal conduction, or unresolved turbulence.}

\begin{figure*}
    \centering
    \includegraphics[width=1.0\textwidth]{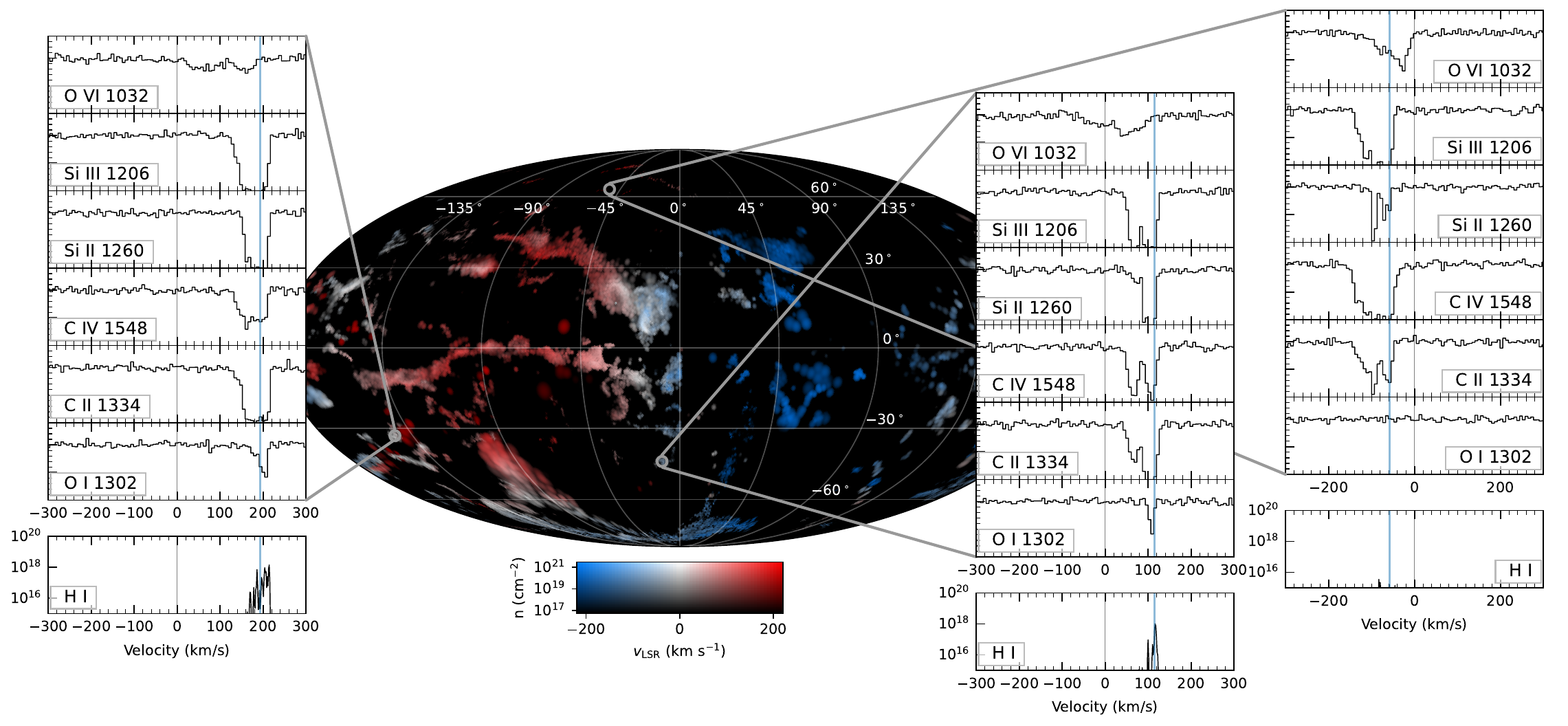}
    \caption{Mock observations of our simulated galaxy. The background image is an all-sky projection from a solar-like position of \ion{H}{1} emission colored by LSR velocity (with brightness denoting column density). The overlaid plots are synthetic spectra towards three of our identified HVC analogs. Each subpanel shows the absorption profile of a different transition as a function of LSR velocity (\rev{absorption from the disk is visible at 0~\kms}{absorption from the disk at 0~\kms\ has been excluded for clarity}). \rev{}{The spectra have resolution of 6.5~\kms, SNR $=25$, and are convolved with a Gaussian LSF for best comparison with STIS E140M spectra.} The blue vertical lines denote the mean LSR velocity of the HVC, and we can see absorption across many different ions aligned with the HVC's neutral gas velocity. The right-most panel shows the spectra for an HVC that is too distant to be observed in \ion{H}{1} emission, however it is still detectable in absorption.}
    \label{fig:spectra}
\end{figure*}

\begin{deluxetable*}{lcccccccc}
\tablecaption{Spectroscopic Total Mass Flow Rates in HVCs}
\label{tab:fc}
\tablehead{ & \multicolumn{4}{c}{Inflowing} & \multicolumn{4}{c}{Outflowing} \\\cmidrule(lr){2-5} \cmidrule(lr){6-9}
    \colhead{} & \colhead{$f_c$} & \colhead{$\langle \log N_\mathrm{H}\rangle$} & \colhead{$\langle v\rangle$} & \colhead{$dM/dt$} & \colhead{$f_c$} & \colhead{$\langle \log N_\mathrm{H}\rangle$} & \colhead{$\langle v\rangle$} & \colhead{$dM/dt$} \\
     &  & (cm$^{-2}$) & (\kms) & (\msolyr) &  & (cm$^{-2}$) & (\kms) & (\msolyr)}

 \startdata
From H (all)  &  $0.34\pm0.05$  &  $19.57\pm0.01$  &  $-86.8\pm0.5$  &  $-1.90\pm0.26$  &  $0.62\pm0.04$  &  $19.43\pm0.02$  &  $+116.5\pm0.4$  &  $3.39\pm0.20$ \\
From H (cool)  &  $0.26\pm0.05$  &  $19.32\pm0.01$  &  $-96.9\pm0.6$  &  $-0.90\pm0.18$  &  $0.14\pm0.05$  &  $19.16\pm0.07$  &  $+89.5\pm1.1$  &  $0.31\pm0.12$ \\
From Si  &  $0.24\pm0.05$  &  $19.49\pm0.01$  &  $-96.3\pm0.7$  &  $-1.25\pm0.26$  &  $0.14\pm0.05$  &  $18.89\pm0.12$  &  $+90.0\pm1.0$  &  $0.17\pm0.06$ \\
\citet{fox19}  &  $0.36\pm0.03$  &  $18.93\pm0.03$  &  $-101\pm10$  &  $-0.53\pm0.23$  &  $0.25\pm0.03$  &  $18.62\pm0.08$  &  $+89\pm12$  &  $0.16\pm0.07$ \\
\enddata

\tablecomments{Columns (1)$-$(4) show the properties of the inflowing gas; columns (5)$-$(8) show the properties of the outflowing gas. Columns (1) and (5) are covering fraction, and the uncertainties are calculated using the Wilson score interval. Columns (4) and (6) are the mean of the log column density. Columns (5) and (7) are the mean velocity, negative denotes inflowing. Columns (6) and (8) are the mass flow rates. The uncertainties in Columns (2)$-$(4) and (6)$-$(8) are calculated via the standard error of the mean.}
\vspace{-0.3in}
\end{deluxetable*}

\rev{}{In terms of our 3D analysis, the structures that we are interested in for this work are comprised of $\sim$100$-$1000+ gas cells which results in good convergence for the study of their global properties. However, we are not able to resolve the details of the mixing layers that exist at the interface between the hot and cold phases, nor the very small cold structures that can form through stochastic variations in densities and temperatures. Specially-designed CGM refinement simulations have shown that with enhanced resolution in these low-density regions, we can expect an increase in the turbulent pressure and velocity dispersion in the hot phase \citep{lochhaas23}, an increase in neutral cold gas mass \citep{vandevoort19}, and more structure at smaller scales \citep{hummels19,augustin25}. Thus, due to the moving-mesh algorithm implemented in Arepo, which mitigates numerical issues with mixing \citep{springel10}, and our focus on large structures that are already well resolved within the simulation, we purport that these derived properties of HVC analogs and their ionized envelopes are quantitative results applicable to the CGM of our own Galaxy.}

\rev{Additionally, TNG50 uses mass-based refinement in which the cell sizes are determined based on their masses.}{That being said, care must needs be taken when analyzing these simulations which rely on mass-based refinement in which the cell sizes are determined based on their masses.} \rev{This leads to larger cells in low-density regions which necessitates care be taken when comparing dense and diffuse regions.}{} \rev{}{Due to the overall density gradient in the gas around the galaxy, this means that the radial position of the clouds within the CGM will affect their resolution. Therefore, we are undertaking ongoing work to not only test these results, but also explore the finer details in small-scale structures and low-density regions by pushing new simulations to the limit of resolution within the CGM.}

\section{Results} \label{sec:results}

\subsection{Mock Observations} \label{sec:mocks}

\rev{We looked at the absorption spectrum for \ion{C}{2}, \ion{C}{4}, \ion{Si}{2}, and \ion{Si}{3} towards each HVC. We also compared against the line of sight \ion{H}{1} column density. Some sample spectra are shown in Figure~\ref{fig:spectra}. This plot shows}{Figure~\ref{fig:spectra} shows} the high-velocity gas in projection on the sky from an assumed solar position at galactocentric radius of 8~kpc (as in \citetalias{lucchini24}). The hue represents velocity in the Local Standard of Rest frame (LSR) and the lightness denotes column density. \rev{The three sets of spectra show the absorptions toward the directions of three HVC analogs.}{We also display the mock spectra for \oi, \cii, \civ, \siii, \siiii, and \ovi\ along with the \hi\ emission along sightlines passing through the cores of three HVC analogs.}

In all cases, we see strong absorption in all ions at the velocities of the HVCs (marked with a blue vertical line). The bottom panel in each set shows the \ion{H}{1} column density in emission, and for two of the HVCs, they are observable in 21~cm emission. However, for the third (right panels), the \ion{H}{1} \rev{}{emission} column densities are too low due to \rev{it being too distant}{its Galactocentric radius being too large}. However, absorption in the other ions \rev{}{(except \oi)} is still observable\rev{}{, due to the increased sensitivity of spectroscopy}.



\rev{}{Following on from these individual sightlines, we used 300 random directions to compute ionic covering fractions and mass flow rates as in \citet{richter17} and \citet{fox19}. First of all, we can compare the covering fractions and mean column densities calculated from silicon using Equation (\ref{eq:nh}), with the directly measured hydrogen columns ($N(\mathrm{H\ I})+N(\mathrm{H\ II})$). However, the silicon is specifically tracing the cool hydrogen, so we can furthermore restrict the $N(\mathrm{H\ I})$ and $N(\mathrm{H\ II})$ calculations to only include gas with temperatures below $10^5$~K. This restriction is denoted as ``cool'' in Table~\ref{tab:fc}.
We use a cutoff column density of $10^{12}$~cm$^{-2}$ for \ion{Si}{2} and \ion{Si}{3}, and a cutoff of $10^{12}/Z(\mathrm{Si}/\mathrm{H})_\odot=8.8\times10^{16}$~cm$^{-2}$ for H (where (Si/H)$_\odot=3.24\times10^{-5}$, and metallicity $Z=0.35$~Z$_\odot$). For the Si to H conversion, we have used metallicities of $Z=0.2$~Z$_\odot$ for the inflowing material, and $Z=0.5$~Z$_\odot$ for the outflowing material, following \citet{fox19}. Our results are listed in Table~\ref{tab:fc}.}

\begin{figure*}
    \centering
    \includegraphics[width=1.0\textwidth]{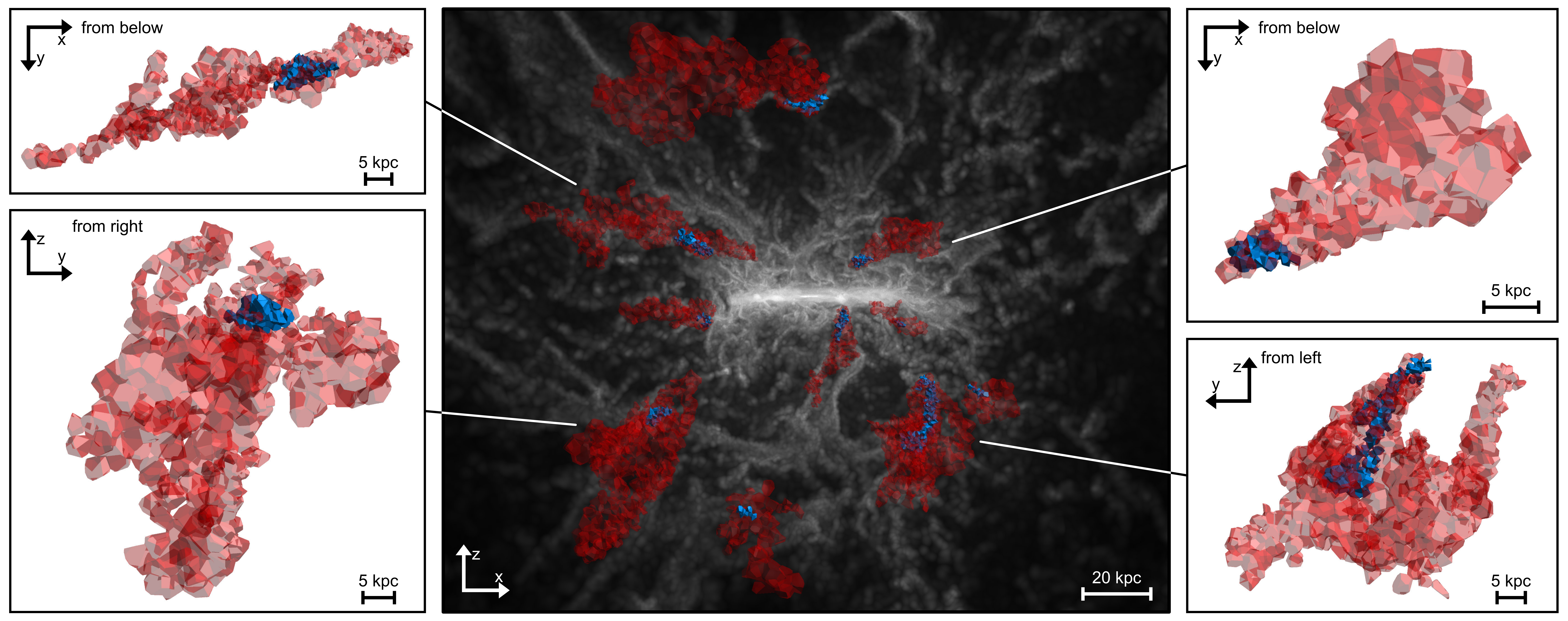}
    \caption{Cartesian view of the galaxy and its HVCs. The central panel shows the gas density in grey in the background with nine 3D HVCs overlaid in blue and red, showing the cool and \rev{}{warm/}hot gas, respectively. These 3D structures are generated by reconstructing the Voronoi mesh and isolating the connected comoving cells (as described in Section~\ref{sec:ion}). Each of the surrounding four panels shows one of the HVCs from an alternate angle.}
    \label{fig:3d}
\end{figure*}

\rev{}{For the inflowing material, we generally see consistent results, both between the H and Si calculations in the simulations, as well as with the observational calculations from \citet{fox19}. However, for the outflowing gas, much of it is warm or hot \citep{clark22}, so the calculation using all the hydrogen gives a significantly higher covering fraction and mean column density than the other methods and the observations. On the other hand, using the cool material in the simulation (either from Si or H), we see a lower covering fraction than what is observed around the MW. Despite lower covering fractions, the mean column densities are slightly higher, so the derived total mass flow rates are still higher than those quoted in \citet{fox19}.}

\subsection{Ionized Envelopes}

\subsubsection{Morphologies}

As described in Section~\ref{sec:ion}, we isolate the ionized envelopes by walking through the Voronoi mesh and identifying connected cells that are comoving with the HVC. The 70 HVC analogs have envelopes ranging from 50 to 1,997 cells. Figure~\ref{fig:3d} shows a selection of these 3D HVCs in projection. In the central panel, the background displays the total projected gas density in grey with nine HVCs (in blue) and their envelopes (in red) overlaid. The outer panels depict four of these HVCs from different angles. The top two are filamentary displaying differing degrees of head-tail structure, while the bottom two are more irregular.

\begin{figure}
    \centering
    \includegraphics[width=0.9\linewidth]{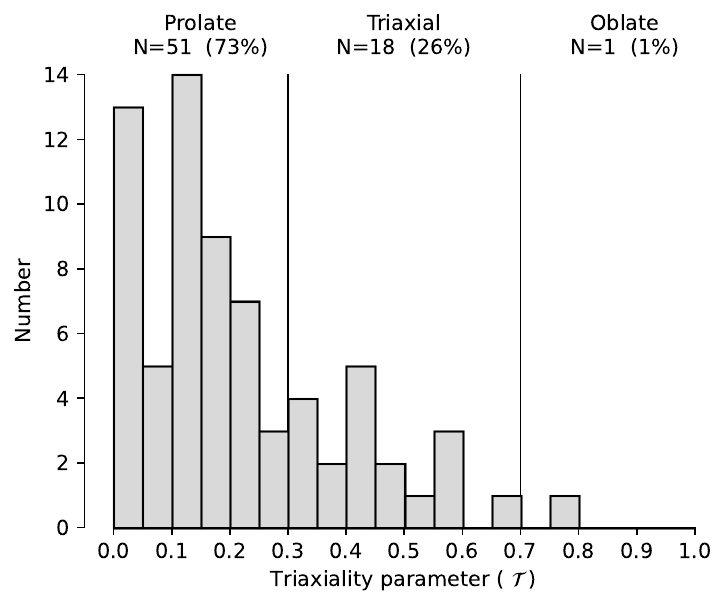}
    \caption{The distribution of the triaxiality parameter, $\mathcal{T}$, defined in Equation~\ref{eq:triax}. Values of $\mathcal{T}\leq0.3$ are considered prolate, $0.3<\mathcal{T}\leq0.7$ are triaxial, and $\mathcal{T}>0.7$ are oblate. We find that the majority of HVC envelopes are prolate, with only a single oblate example.}
    \label{fig:triax}
\end{figure}

To quantify this large variety in morphologies, we use the ratios of their principal axes. For each HVC envelope, we calculate its moment of inertia tensor, $I_{ij}=\sum_{k=1}^n m_k \left(r_k^2\delta_{ij}-x_i^{(k)}x_j^{(k)}\right)$, where $i,j$ can be 1, 2, or 3 (corresponding to $x$, $y$, and $z$), and the sum is over all the particles denoted by $k$. $x_i^{(k)}$ is the $x$, $y$, or $z$ position of particle $k$ within the envelope, and $m_k$ is its mass. We then determine its sorted eigenvalues, $a$, $b$, and $c$ (s.t. $a<b<c$) and calculate axis ratios $p=\sqrt{a+b-c}/\sqrt{-a+b+c}$ and $q=\sqrt{a-b+c}/\sqrt{-a+b+c}$ (following the convention that $p<q<1$). These values are then combined into a ``triaxiality parameter'' defined as
\begin{equation} \label{eq:triax}
    \mathcal{T}=\frac{q^2-p^2}{1-p^2}.
\end{equation}
We classify an envelope to be prolate for values of $\mathcal{T}\leq0.3$, triaxial for $0.3>\mathcal{T}\geq0.7$, and oblate for $\mathcal{T}>0.7$. Figure~\ref{fig:triax} shows the distribution of triaxialities for our 70 HVC envelopes. We find that majority to be prolate (51/70, 73\%), consistent with the picture that these objects are associated with filamentary structures. The subdominant population are triaxial (18/70, 26\%), and we find only a single oblate envelope.

With the majority of the HVC envelopes being prolate and filamentary, we also investigated their head-tail structure. We determine whether an HVC and its envelope exhibit a head-tail structure by looking at the separation between the HVC's center of mass and the center of mass of its envelope normalized by the maximum radial extent of the envelope, defined as $f_\mathrm{ht}=\vert \bar{\mathbf{r}}_\mathrm{HVC}- \bar{\mathbf{r}}_\mathrm{ion} \vert / \max(\vert \bar{\mathbf{r}}_\mathrm{ion}-\mathbf{r}_i \vert)$. If the HVC is at the edge of the envelope we find $f_\mathrm{ht}\approx1$, if it is at the center, $f_\mathrm{ht}\approx 0$. Based on visual inspection, $f_\mathrm{ht}>0.5$ constitute clouds that appear to have a head-tail structure, and this includes $\sim$10\% of the HVCs (with slight variations depending on whether we cut off the morphologies at 1,000, 2,000, or 3,000 cells).

\subsubsection{Masses \& Accretion Rates} \label{sec:masses}

We calculate the total mass in the HVCs' ionized envelopes as discussed above (Section~\ref{sec:ion}) by summing the \rev{}{warm and} hot gas ($>10^{4.5}$~K) within two half-sizes weighted by the alignment factor, $dv_i$. Building on \citetalias{lucchini24}, we can investigate these ionized envelopes as a function of formation pathway of the HVC. In that work, we categorized each HVC based on its history into having originated from the galactic disk, from stripped satellite material, or from thermal instability (TI) of warm or hot material. In this paper, we combine the TI categories for simplicity.

\begin{figure}
    \centering
    \includegraphics[width=0.8\linewidth]{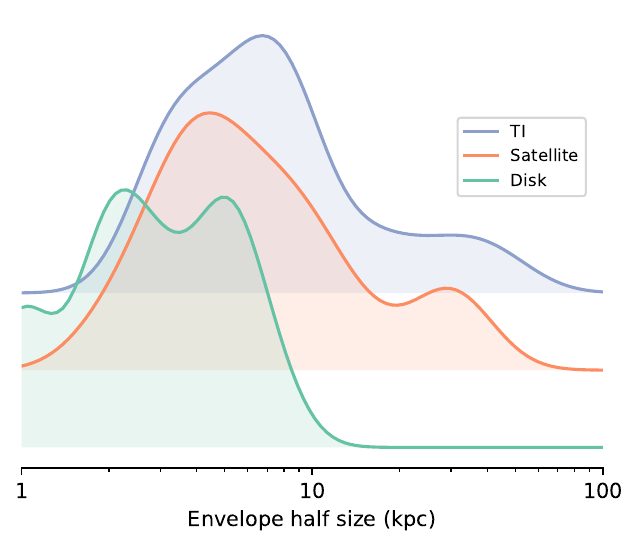}
    \caption{Distributions of calculated half-sizes separated by HVC origin as determined in \citetalias{lucchini24}. We see that the disk origin HVCs tend to have smaller half-sizes, while the stripped satellite material and thermal instability clouds tend to coexist with larger ionized envelopes.}
    \label{fig:sizes}
\end{figure}

\begin{figure}
    \centering
    \includegraphics[width=1.0\linewidth]{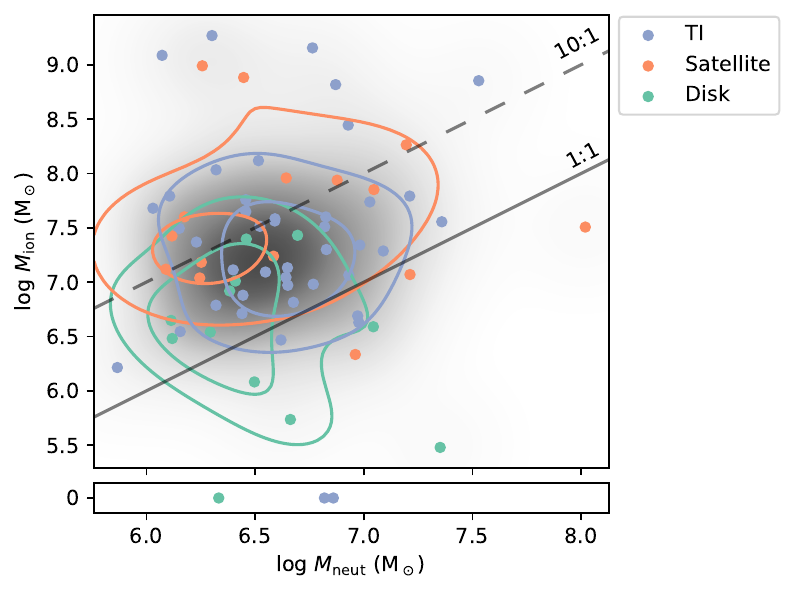}
    \caption{The total ionized mass vs. neutral mass for each HVC analog in our galaxy. The grey background shows the distribution for all clouds, while the colored dots and contours show the populations separated by origin as determined in \citetalias{lucchini24}. The lines mark where the ionized mass is equal to the neutral mass (solid, 1:1), or where the ionized gas is ten times greater (dashed, 10:1). The mean ratio of ionized to neutral mass is 5.6.}
    \label{fig:mass_ratios}
\end{figure}

Figure~\ref{fig:sizes} shows the distribution of half-sizes separated by origin$-$thermal instability in blue, stripped satellite material in orange, and disk origin in green. We see an increasing median half-size as we go from disk, to satellite, to TI origins. This can also be seen in Figure~\ref{fig:mass_ratios} which shows the distribution of ionized envelope mass vs. cold HVC mass separated by origin. The background greyscale shows the total distribution, while the colored points and contours show the distributions for each formation pathway. The solid and dashed black lines also denote the one-to-one and ten-to-one ratios, respectively. The mean ratio of hot to cool mass is 5.6, i.e. there is $\approx5.6$ times more hot mass than cold mass for a given HVC on average. Furthermore, disk origin HVCs tend to have less material in their ionized envelopes while satellite origin and thermal instability origin HVCs have a broader spread.

As described above (Section~\ref{sec:ion}), we calculate the accretion rates due to the neutral HVCs as well as their ionized envelopes. We find a rate of 1.3 \msolyr\ for the HVCs, and a rate of 6.8 \msolyr\ for the ionized envelopes. The star formation rate for this galaxy (calculated by SUBFIND) is 5.8 \msolyr. This \rev{confirms}{aligns with} the observational estimates from our own Galactic HVCs stating that neutral HVCs can account for $\sim$10\% of Galactic star formation, while the ionized material falling onto our galaxy through the CGM could provide the dominant contribution to sustain the Milky Way's SFR \citep{fox19}. In this simulated galaxy, we find that neutral HVCs can sustain 22\% of the SFR, while the ionized envelopes have a larger accretion rate of 117\% of the SFR. Below we extend this analysis to 47 other MW-like galaxies in TNG50.

\begin{figure*}
    \centering
    \includegraphics[width=1.0\textwidth]{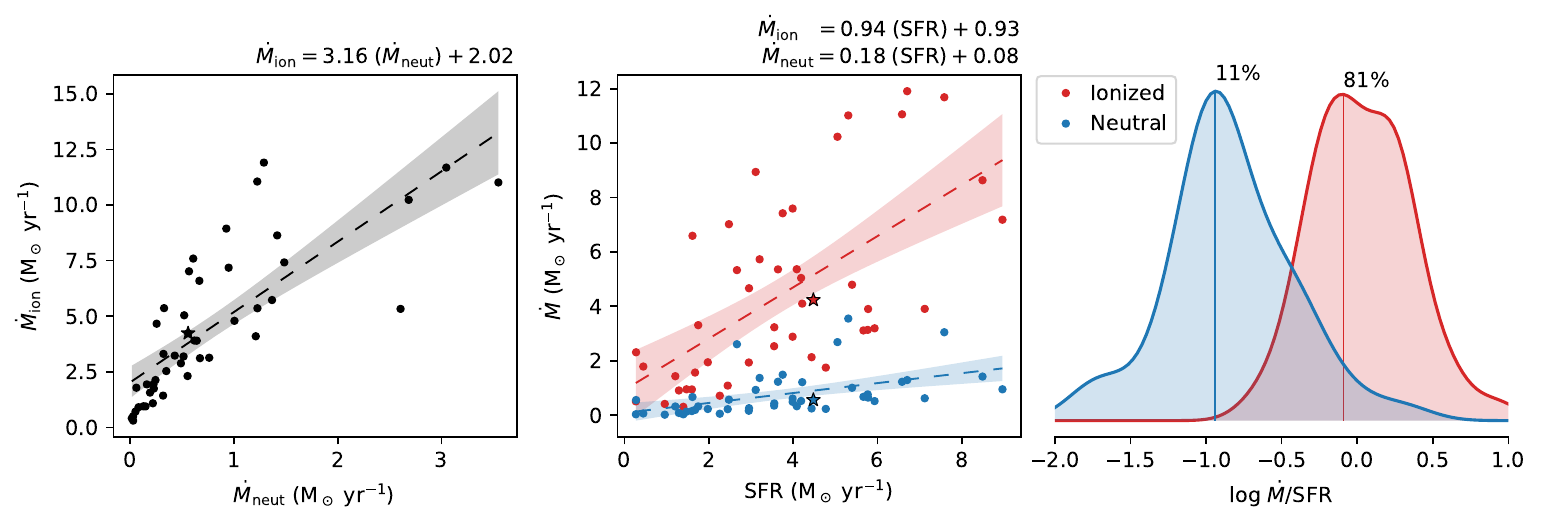}
    \caption{Accretion rate values for the population of MW-like galaxies from \citet{semenov24}. \textit{Left}: The calculated accretion rate of the ionized HVC material vs. the neutral HVC gas. Each point represents the summed total accretion for all HVCs within a single galaxy. A linear fit to the data is also shown with equation $\dot{M}_\mathrm{ion}=3.16\;(\dot{M}_\mathrm{HVCs})+2.02$. \textit{Center}: Accretion rates for ionized (red) and neutral (blue) HVCs as a function of the galaxy's star formation rate. Linear fits again are shown as dashed lines with equations $\dot{M}_\mathrm{ion}=0.94\;(\mathrm{SFR})+0.93$ and $\dot{M}_\mathrm{HVCs}=0.18\;(\mathrm{SFR})+0.08$. \textit{Right:} Distributions of accretion rate divded by star formation rate for the ionized (red) and neutral (blue) material. The peaks, located at 11\% and 81\%, are denoted with vertical lines and labeled.}
    \label{fig:trio}
\end{figure*}

\subsection{Population Analysis} \label{sec:population}

We also explored the HVC population around other MW-like galaxies in TNG50. We used the sample from \citet{semenov24} selected with more stringent criteria than the MW/M31 sample available as part of the TNG50 data release \citep{pillepich24}. Galaxies were selected based on total mass ($M_{200}=0.8-1.4\times10^{12}$~\msol), star formation rate (SFR, $\dot{M}_*>0.2$~\msolyr), and diskiness ($(v_\mathrm{rot}/\sigma)>6$ for $<100$~Myr old stars). See \citet{semenov24} for more details. Our sample here consists of the 52 galaxies that overlap between the \citet{semenov24} sample and the TNG50 MW/M31 sample due to the availability of 800~kpc cube cutouts for the MW/M31 galaxies. Of these 52, 6 were excluded as they are undergoing major mergers at $z=0$ (excluded IDs: 508538, 511303, 526029, 527309, 530852, 542662).

In these 46 galaxies, we followed the steps described in \citetalias{lucchini24} to identify contiguous cold clouds and compare their velocities with projected disk velocities based on the galactic rotation curves. Thus for each galaxy, we have a sample of HVC analogs and we can characterize their properties as well as the properties of their ionized envelopes. We calculate the accretion rates due to the HVCs as well as due to their ionized gas envelopes (defined using the alignment fraction, see Section~\ref{sec:ion}). We then compare these values with the SFR calculated via the SUBFIND algorithm run as a part of the TNG50 simulation and accessible via their online API\footnote{\url{https://www.tng-project.org/data/}}.

As we found with our fiducial galaxy above (537941), the total infalling material associated with HVCs is dominated by their ionized envelopes. Figure~\ref{fig:trio} contains our main results. The left panel shows the total accretion rate of the ionized envelopes vs. that of the neutral HVCs for each galaxy. The points are fit with a line, \rev{$\dot{M}_\mathrm{HVC}=2.9\times\dot{M}_\mathrm{ion}+2.1$}{$\dot{M}_\mathrm{neut}=3.16\times\dot{M}_\mathrm{ion}+2.02$}, showing that the accretion rate from the ionized material is $\sim$3 times more than the neutral accretion.

Furthermore, the total accretion from HVCs and their ionized envelopes scales with SFR. The middle panel of Figure~\ref{fig:trio} shows the scaling of the accretion rate for the ionized gas (red) and neutral gas (blue) with SFR. The slopes of the fit lines are $0.94\pm0.18$ and $0.18\pm0.05$ for the ionized and neutral rates, respectively. This can be interpreted as the accretion of the ionized gas being able to balance 94\% of the galactic SFR, while the neutral material can only account for 18\%.

This can been seen more specifically on a galaxy-by-galaxy basis in the right panel. Here, we are plotting the distribution of accretion rate to SFR ratios for the neutral gas in blue, and the ionized gas in red. The vertical lines denote the peak of the distributions at 11\% and 81\% for the neutral and ionized material, respectively. As deduced by the slopes of the line fits in the middle panel of Figure~\ref{fig:trio}, we see that the accretion of the ionized envelopes around the HVCs can balance the galactic star formation rate, while the neutral HVC material can only account for a small fraction. \rev{}{Combining the neutral and ionized material into a total accretion rate (as is done in the observations), the $\dot{M}/$SFR distribution is peaked at 93\%.}

\section{Discussion} \label{sec:discussion}

To calculate the ionized gas masses and accretion rates, we summed all the aligned material within twice the calculated half-sizes for each HVC (Section~\ref{sec:ion}, Figure~\ref{fig:alignfrac}). However, we also wanted to compare these values to the results directly using the full 3D envelopes. For this method, we summed all the gas with $T>10^{4.5}$~K within the 3D envelopes. This gave similar results with 80\% of the HVCs in our primary galaxy having 3D envelope masses within 15\% of the masses calculated with the half-size technique. The few outliers were envelopes that changed dramatically based on the maximum envelope size that we imposed (2,000 cells).

While the correspondence between ionized HVC accretion rate and SFR may be coincidence, it is helpful to compare the high-velocity portion of accretion with that of other CGM populations. The accretion rate due to all the infalling material \textit{above} $10^{4.5}$~K is, on average, $\sim7$ times more than that of only the material comoving with the HVC analogs. \rev{While}{On the other hand,} the accretion rate from all the infalling material \textit{below} $10^{4.5}$~K is $\sim25$ times more than that of the identified \rev{cold clouds}{HVC analogs themselves}. Future work is required to determine what the actual source of star forming material is for these galaxies, however the fact that the high-velocity hot material represents a larger portion of the total indicates that this is perhaps the dominant contribution. Additionally, it has been shown that in order to replenish molecular gas for sustained star formation, ionized material is required \citep{bauermeister10}. Replenishment from \ion{H}{1} is not sufficient due to the minimal evolution of \ion{H}{1} densities from $z=4$ \rev{}{\citep{peroux20}}.

However, in order to eventually form stars, the ionized material would need to cool and condense enough to form neutral and molecular gas that could collapse into stars. The HVC envelopes are infalling (by selection) which means that they will continue to encounter denser material which would lead to them becoming underpressured and thus collapsing. Additionally, clouds could be broken up due to increased galactic shear as they approach the disk, leading to smaller, denser clouds close to the disk. While we do see a positive trend between HVC envelope size and galactocentric radius in our simulations (smaller envelopes at smaller radii), this may be a resolution effect since the gas cell sizes are smaller at smaller radii due to the overall density of the CGM being higher. \ion{H}{1} cloud size distributions have been studied in the past with enhanced refinement schemes \citep{hummels19,gible}, however future applications to the\rev{ sizes of}{} ionized high-velocity envelopes are required. \rev{Additionally, future work explicitly exploring the future of these clouds will illuminate how their sizes and densities change as they move through the CGM.}{}

\rev{}{We also computed the total mass flow rates of the infalling material via mock spectroscopy to better compare against observational studies. In Table~\ref{tab:fc}, we show that while the simulation generally underpredicts the total amount of gas in the CGM, it is at a higher column density and so the calculated mass flow rates are still higher than those computed from observed spectra. However, this is only at the 1$-$2$\sigma$ level when comparing the observations with the cool ions (calculated via \ion{Si}{2} and \ion{Si}{3}, or via restricting the total hydrogen based on temperature; see Section~\ref{sec:mocks}). The covering fractions and column densities when including the hydrogen at all temperatures in the simulation are larger, especially for the outflowing material. This is due to the fact that the outflows from the galaxy are largely the hot remnants of supernovae which would mostly be traced by higher ions.}

\rev{}{Interestingly, even using the total H mass flow rate from the 300 random spectra, we substantially underestimate the amount of infalling gas onto the galaxy. The mock spectra indicate $1.90\pm0.26$~\msolyr, while the actual rate calculated from the neutral and ionized HVCs is $8.2$~\msolyr. We believe that is mainly due to the fact that a distance must be assumed to turn the column density into a total mass. In this work, we used $d=12$~kpc as done in \citet{fox19}, which is based on observational constraints within the MW. However the gas distances in the simulation are greatly variable (and generally larger than observed within the MW) and can result in vastly different accretion rates. If we simply sum up the accretion rates from all the individual gas cells, we find 65~\msolyr.}

\rev{}{Small-scale cloud crushing simulations have provided an avenue to explore the evolution of HVC-like objects in the CGM as well. Early work showed the importance of radiative cooling in cloud survival through mixing at the turbulent interface between the cold cloud and its warm/hot surroundings \citep{gronke18,gronke20,li20,fielding20}. Recent studies have included stratified background densities (as you would see in the CGM), turbulence, magnetic fields, and thermal conduction \citep{gronke22,tan23,kaul25,yang25}. Clouds are still able to survive in turbulent environments \citep{gronke22}; depending on the orientation, magnetic fields can enhance survival \citep{kaul25}; and thermal conduction seems to inhibit cloud survival \citep{yang25}. However the inclusion of a stratified medium gets us closer to answering the question of whether these clouds can survive long enough to directly interact with the disk. \citet{tan23} still find cloud growth and survival in addition to terminal velocities due to accretion drag, albeit modified for the stratified background scenario. We postpone the analysis of the fate of our HVC analogs within the IllustrisTNG simulations to future work due to the limitations of the physical model of the interstellar medium in TNG.}

Around our own Galaxy, the Magellanic Clouds loom large and have the potential to dramatically impact the Milky Ways' future evolution \citep{fox14,donghia16,lucchini20,lucchini24c}. In our sample of MW-analogs, we excluded several with strong interactions at the present day, however in the galaxy studied in \citetalias{lucchini24} (537941), there were several massive HVCs comprised of gaseous material that had been stripped out of a satellite galaxy. This satellite is in a much more advanced stage of interaction with the central, so it is difficult to compare directly, but we did find that the HVCs associated with the satellite material are larger and contain more neutral and ionized material, in agreement with our observations of the Magellanic System. Working with a more comprehensive simulation of the full Local Group could provide further insights on this front in the future.

We also explore the properties of the ionized envelopes themselves. \rev{}{What surrounds HVCs has long been of interest in uncovering the properties of the MW's broader CGM and can help us understand the pressure equilibrium (or lack thereof) of these multiphase objects \citep{benjamin04}.} Their temperature distributions peak \rev{}{slightly} above the average temperature of the hot phase of the CGM\rev{ ($10^{5.8}$ vs $10^{5.5}$~K)}{} indicating that the ionized envelopes of HVC analogs are in general hotter than the surrounding diffuse gas. \rev{}{Closest to the neutral clouds, the envelopes are still cool/warm, however their temperature increases rapidly with increasing distance from the cloud, peaking a few kpc away and then dropping to the mean ambient temperature. The ambient CGM has a peak in its temperature distribution at $3.3\times10^5$~K with a 16$-$84\% spread of $1.4-5.4\times10^{5}$~K. In order to quantify the temperatures of the ionized envelopes, we took the peak temperature value in the distributions of each envelope's cells ($\sim$ the mode of the distribution), then looked at the distribution of those peak temperatures. The 16$-$84\% range for this distribution is $3.4-7.4\times10^5$~K.} This could be due to the fact that these clouds are selected based on their high anomalous velocities, and thus they will be experiencing increased shear from the surrounding medium which could increase their temperatures.


The ionized HVC envelopes are mostly prolate which indicates that they are filamentary in structure. In the ISM, it has been shown that linear, filamentary structures dominate and are strongly aligned with the magnetic field \citep{clark19}, however we don't yet have corresponding observations in the CGM. We may expect filamentary structures in the CGM when strong magnetic fields are present, when the material is dense enough to collapse under self-gravity, or when there are strong velocity shears. In this case, we expect that the prolate nature of our ionized HVCs is a selection effect based on their high deviation velocities which leads to strong velocity shear with respect to the ambient CGM. However, a further investigation of the morphologies of CGM structures is warranted. The effect of magnetic fields on the morphologies of neutral gas clouds in the CGM has been performed with the high-resolution GIBLE simulations \citep{ramesh24Bfields}, and TNG50 also include magnetic fields, so the B field structure of our ionized HVC envelopes identified here will be explored in future work.

The formation of these ionized envelopes has also been richly studied. They could become ionized through conductive interfaces \citep{borkowski90,gnat10}, turbulent mixing layers \citep{slavin93,kwak10,fielding20,tan21}, shock ionization \citep{bland-hawthorn07,tepper-garcia15}, and/or photoionization \citep{bland-hawthorn13,putman03}. The TNG50 simulations can track shocks well due to the moving-mesh Voronoi tesselation hydrodynamics scheme \citep{springel10}, and include photoionization from a redshift-dependent UV background \citep{nelson19b,faucher-giguere09}. However, thermal conduction is not explicitly included and the cell sizes needed to resolve the turbulent mixing layers are much smaller than those of TNG50. Small-scale ``cloud crushing'' simulations have begun to explore these objects indicating that conduction and photoionization may play a significant role in their morphology and ion structure \citep{yang25}. Further simulations are required to fully explore the formation of these HVCs and their multiphase structure.

\section{Conclusions} \label{sec:conclusions}

In this work we have found that the ionized component of the high-velocity sky in simulated galaxies dominates in mass and accretion rate over the neutral material visible in \ion{H}{1} emission. Additionally, the accretion rate of the ionized HVC envelopes is comparable to the galaxies' star formation rates, showing that these objects play an important role in galaxy growth and the sustaining of star formation. This is consistent with previous estimates for our own Galaxy \citep{lehner11,fox19}. Furthermore, we have characterized the morphologies and properties of the ionized HVC envelopes finding that they are mostly prolate \rev{}{and} generally hotter than the ambient diffuse CGM gas\rev{, and intermediate metallicity between the CGM's cool and hot phases}{}.

In the future, we will characterize the eventual fate of this material to determine the true source of the star forming material in MW-like galaxies. Furthermore, with improved resolution we will be able to investigate \rev{}{more realistic spectra, detailed ionic covering fractions, and} the mixing and interactions between the multiple phases of the CGM in significantly improved detail.

\pagebreak
    The authors thank the referee for their productive comments to improve the manuscript.
    SL thanks Trey Wenger and Cameron Hummels for productive discussions throughout the development of this work. Support for SL was provided by Harvard University through the Institute for Theory and Computation Fellowship. The computations in this paper were run on the FASRC cluster supported by the FAS Division of Science Research Computing Group at Harvard University.

\bibliography{references}{}
\bibliographystyle{aasjournal}

\end{document}